\documentclass{emulateapj}

\usepackage{lscape}
\usepackage{rotate}
\usepackage{longtable}
\usepackage{natbib}

\shorttitle{Chandra Observations of the Cl1604 Supercluster}
\shortauthors{Kocevski et al.}

\begin{document}

\title{\emph{Chandra} Observations of the Cl1604 Supercluster at z=0.9: Evidence for an Overdensity of Active Galactic Nuclei}
\author{Dale D. Kocevski, Lori M. Lubin, Roy Gal\altaffilmark{1}, Brian C. Lemaux, Christopher D. Fassnacht, Gordon K. Squires\altaffilmark{2}}

\affil{Department of Physics, University of California, Davis, 1 Shields Avenue,  Davis, CA 95616}
\altaffiltext{1}{Institute for Astronomy, University of Hawaii, 2680 Woodlawn Dr., Honolulu, HI 96822}
\altaffiltext{2}{\emph{Spitzer} Science Center, M/S 220-6, California 
                 Institute of Technology, 1200 East California Blvd, Pasadena, CA, 91125}
\email{kocevski@physics.ucdavis.edu}

\begin{abstract}
We present the results of \emph{Chandra} observations of the Cl1604
supercluster at $z\sim0.9$.  The system is the largest structure mapped
at redshifts approaching unity, containing at least eight spectroscopically
confirmed galaxy clusters and groups.  Using two 50-ksec ACIS-I pointings we
examine both the X-ray point source population and the diffuse emission from
individual clusters in the system. We find a $2.5\sigma$ excess of point
sources detected in the hard band (2-10 keV) relative to the number of sources
found in blank fields observed by \emph{Chandra}. No such excess is observed
in the soft band (0.5-2 keV).  The hard-band source density is 1.47 times
greater than that of a blank field, in agreement with the previously reported
correlation between overdensity amplitude and cluster redshift. Using a
maximum likelihood technique we have matched 112 of the 161 detected X-ray
point sources to optical counterparts and found 15 sources that are
associated with the supercluster.  All 15 sources have rest-frame
luminosities consistent with emission from active galactic nuclei (AGN).  We
find that the supercluster AGN largely avoid the densest regions of the
system and are instead distributed on the outskirts of massive clusters or
within poorer clusters and groups.  We have also detected diffuse emission
from two of the eight clusters and groups in the system, clusters
Cl1604+4304 and Cl1604+4314.  The systems have bolometric luminosities of
$1.43\times10^{44}$ and $8.20\times10^{43}$ $h_{70}^{-2}$ erg s$^{-1}$ and
gas temperatures of $3.50^{+1.82}_{-1.08}$ and $1.64^{+0.65}_{-0.45}$ keV,
respectively.  Using updated velocity dispersions, we compare the properties
of these systems to the cluster scaling relations followed by other X-ray and
optically selected galaxy clusters at high redshift. 
\end{abstract}


\section{Introduction}

There is an ongoing debate as to whether active galactic nuclei (AGN) are
triggered in the environments around galaxy clusters in excess to what is
observed in the field.  A growing number of studies have reported an
overdensity of X-ray point sources in the vicinity of clusters relative to
blank-field observations (Henry \& Briel 1991;  Cappi et al.~2001; Pentericci
et al.~2002; Molnar et al.~2002, Johnson et al.~2003; D'Elia et al.~2004; Ruderman \& Ebeling
2005; Cappelluti et al.~2005; Hudaverdi et al.~2006; Branchesi et al.~2007),
while other studies like the large \emph{Chandra} archival program ChaMP (Kim
et al.~2004a), using observations covering 1.1 deg$^{2}$, find no significant
difference between cluster and cluster-free fields (Molnar et al.~2002; Kim et al.~2004b).

Many studies which have reported overdensities find evidence that the
excess sources are AGN on the outskirts of clusters and that they may
trace the large-scale structure which surrounds the systems.  D'Elia et al.~(2004)
found an asymmetric distribution of point sources around the  $z=0.46$
cluster 3C 295, which the authors propose may trace a large-scale filament
connected to the cluster.  Johnson et al.~(2003) report a $2\sigma$
overdensity of AGN in MS1054-0321 and found that the sources were
preferentially located 1-2 Mpc from the cluster center.  Likewise, Ruderman
\& Ebeling (2005) reported a statistical excess of AGN near the virial radius
of clusters after combining the source counts of 51 high-redshift clusters
from the MACS survey (Ebeling et al.~2001).  

There is also evidence that the amplitude of the observed source overdensity
increases with redshift.  Cappelluti et al.~(2005) recently performed the
first systematic search for X-ray point source overdensities in the fields of
10 high-redshift ($0.24<z<1.2$) clusters observed with \emph{Chandra}.  They
found that $40\%$ of the cluster fields showed an excess of point sources and noted a
correlation between the amplitude of the overdensity in the hard band and cluster redshift.
A similar conclusion was reached by Branchesi et al.~(2007) who examined the fields of 18 distant
clusters ($0.25<z<1.1$).  Such a correlation would be expected if AGN are
indeed tracing the cosmic web near clusters, since numerical simulations tell us that
a greater degree of large-scale structure should exist around dynamically younger
clusters at high redshift (Colberg et al.~2000; Evrard et al.~2002).

The findings that AGN may trace the filaments which feed clusters is
consistent with studies that suggest the large-scale
structure surrounding clusters plays a pivotal role in driving
galaxy evolution.  Studies at low redshift have found that regions of intermediate density, such
as the groups and filaments on the outskirts of clusters, already exhibit
signs of galaxy transformation and reduced star formation rates (Lewis et
al. 2002, Gomez et al.~2003).  Many of the processes which have been proposed
to drive this evolution, such as galaxy mergers (Barnes \& Hernquist 1991)
and galaxy harassment (Moore et al.~1996), may also work to funnel gas to the
central regions of galaxies initiating AGN activity.  

 Thus far it has been difficult to conclusively establish whether AGN are triggered near
 clusters and to determine the processes which may be responsible because the
 observed overdensities are relatively mild, resulting in a limited number of
 confirmed, optically identified cluster AGN available for study in a single
 system.  Superclusters offer an alternative to studying individual clusters
 or the AGN population of a statistical sample of clusters.  These
 large-scale systems are comprised of several galaxy clusters and groups
 connected by a rich network of filamentary structure on scales of 10-100 Mpc
 (Bahcall \& Soneira 1984; Einasto et al.~2001), thus potentially providing
 large samples of AGN at the same epoch.  Such systems not only provide
 significant large-scale structure within which AGN may be preferentially
 found, but also a wide variety of environments and local conditions to help
 constrain the mechanisms most responsible for triggering their activity. 

Recently Gilmour et al.~(2007) used \emph{XMM-Newton} observations of the Abell 901/902
supercluster at $z=0.17$ to examine the environments and optical properties of
X-ray selected AGN in the system.  The authors find that AGN are more
prevalent in clusters than optical studies have suggested, in agreement with
the results of Martini et al.~(2002, 2006).  They also note that AGN host galaxies
are preferentially found in areas of modest galaxy density and strongly avoid
the densest regions of the supercluster, concluding that there are strong
correlations between AGN activity and local environment.  Furthermore they find that
the local densities and optical colors of the AGN host galaxies are more
comparable to galaxy groups and the outskirts of clusters than filament and
cluster-like environments.  

While low-redshift superclusters like Abell 901/902 have been well cataloged (Bahcall \& Soneira
1984; Tully et al.~1992; Einasto et al.~2001) and studied for some time
(e.g.~Shapley 1930), only a limited number of such systems
are known at higher redshifts.  These include a structure at $z=0.89$ detected in the UK Infrared Deep
Survey (UKIDSS; Swinbank et al.~2007), the RCS2319+00 supercluster of three
X-ray luminous clusters at $z=0.9$ (Gilbank et al.~2008), and the Lynx system
at $z = 1.27$ which contains at least two clusters (Nakata et al.~2005).

In this paper we report on \emph{Chandra} observations of the Cl1604
supercluster at $z=0.9$.  The system is the largest supercluster mapped at
redshifts approaching unity, with the most constituent clusters and groups and the
largest number of spectroscopically confirmed member galaxies.  We here make
use of two 50 ksec pointings to examine both the diffuse emission from
the system's clusters and the properties of point sources in the two fields.

Originally detected as two rich clusters in the optical survey of Gunn et al.~(1986),
follow-up wide field imaging of the Cl1604 system revealed 10 distinct red galaxy
overdensities within a $25^{\prime}\times25^{\prime}$ region on the sky ( Lubin et al.~2000;
Gal \& Lubin 2004).  Extensive spectroscopic observations confirmed that four
of these overdensities were galaxy clusters with velocity dispersion greater
than 500 km s$^{-1}$, while an additional four were found to be poorer clusters
and/or groups ($\sigma = 300-500$ km s$^{-1}$; Postman et al.~1998, 2001; Gal
et al.~2005, 2008).  Thus far over 1400 extragalactic redshifts have been
compiled in the field of the Cl1604 system, resulting in spectra for 449 confirmed
supercluster members (Gal et al.~2008).  

The richest cluster in the system, Cl1604+4304, was previously observed with
\emph{XMM-Newton} by Lubin et al.~(2004), who found the system to have a
bolometric luminosity of $L^{\rm bol}_{\rm x} = 2.01\times10^{44}$ $h_{70}^{-2}$ erg s$^{-1}$ and
temperature of $T=2.51^{+1.05}_{-0.69}$ keV.  The second cluster originally
detected by Gunn et al.~(1986), Cl1604+4321, was not detected in \emph{ROSAT}
observations of the system and has an $3\sigma$ luminosity upper limit in the
0.1-2.4 keV band of $L_{\rm x} \le 4.76 \times 10^{43}$ $h_{70}^{-2}$ erg
s$^{-1}$ (Castander et al.~1994; Postman et al.~2001).  The remaining six newly discovered 
clusters and groups in the supercluster have not been previously observed at
X-ray wavelengths.  Five of these systems fall within the field-of-view (FOV)
of the \emph{Chandra} observations presented here.

This study of the Cl1604 supercluster is organized as follows:  In \S\ref{sect-data} we describe the
X-ray observations and our data reduction procedures, as well as a summary of the
optical imaging and the extensive spectroscopic dataset available for the
system.  \S\ref{sect-xps} discusses the properties of the detected X-ray point sources,
including their logN-logS distribution, optical counterparts and redshift
distribution.  In \S\ref{sect-diff} we present the X-ray luminosities and gas
temperatures of the diffuse cluster emission detected in the system, followed by a comparison
of the cluster X-ray properties to cluster scaling relations.  Throughout
this paper we assume a $\Lambda$CDM cosmology with $\Omega_{m} = 0.3$, $\Omega_{\Lambda}
= 0.7$, and $H_{0} = 70$ $h_{70}$ km s$^{-1}$ Mpc$^{-1}$.

\section{Observations and Data Reduction}
\label{sect-data}

\subsection{X-ray Observations}

 Observations of the Cl1604 supercluster were carried out with {\it
   Chandra's} Advanced CCD Imaging Spectrometer (ACIS; Garmire et al.~2003) on 2006
 June 23 (obsID 7343), June 25 (obsID 6933), and October 01 (obsID 6932).  A
 summary of the observational parameters of the three datasets are listed in
 Table \ref{tab_obspar}.  The three observations consist of two pointings,
 one encompassing the northern portion of the system and the other the
 southern portion, with a $4\farcm9$ overlap between the imaged regions.
 The aimpoints of the observations are $\alpha_{2000} = 16^{\rm h}04^{\rm
   m}12.0^{\rm s}$, $\delta_{2000} = +43^{\circ}22^{\prime}35^{\prime\prime}$
 and $\alpha_{2000} = 16^{\rm h}04^{\rm m}19.5^{\rm s}$, $\delta_{2000} =
 +43^{\circ}10^{\prime}31^{\prime\prime}$.   Imaging of the northern pointing
 was split between two observations (obsID 7343 and 6933) with nearly
 identical aimpoints and roll angles, while the southern pointing was covered
 by a single observation (obsID 6932).  Each pointing was imaged with the
 $16\farcm9\times16\farcm9$ ACIS-I array, with the aimpoint located on the
 ACIS-I3 chip.  The ACIS-S2 chip was also active during the observations but
 due to its large off-axis angle and reduced effective area we do not make
 use of it in this analysis. All three observations were carried out in VFAINT
 telemetry mode with the nominal 3.2 sec CCD  frame time for a total
 integration of 19.4, 26.7, and 49.5 ksec for the 7343, 6933, and 6932
 datasets, respectively.  An examination of light curves produced in the
 0.3-10 keV band shows no indication of flaring during the course  of the observations.

\begin{center}
\begin{deluxetable}{lcccc}
\tablewidth{0pt}
\tablecaption{Summary of Observations \label{tab_obspar}}
\tablecolumns{5}
\tablehead{\colhead{Target} & \colhead{Obs ID} & \colhead{Exp (s)}  & \colhead{Obs Date} & \colhead{Detector}}
\startdata
Cl1604-North &  6933  &  26691  &  2006 Jun 25  &  ACIS-I0123  \\ 
Cl1604-North &  7343  &  19412  &  2006 Jun 23  &  ACIS-I0123  \\ 
Cl1604-South &  6932  &  49478  &  2006 Oct 01  &  ACIS-I0123  \\
\vspace*{-0.075in}
\enddata
\end{deluxetable}
\end{center}
\vspace*{-0.45in}

\begin{figure*}
\epsscale{1.0}
\plotone{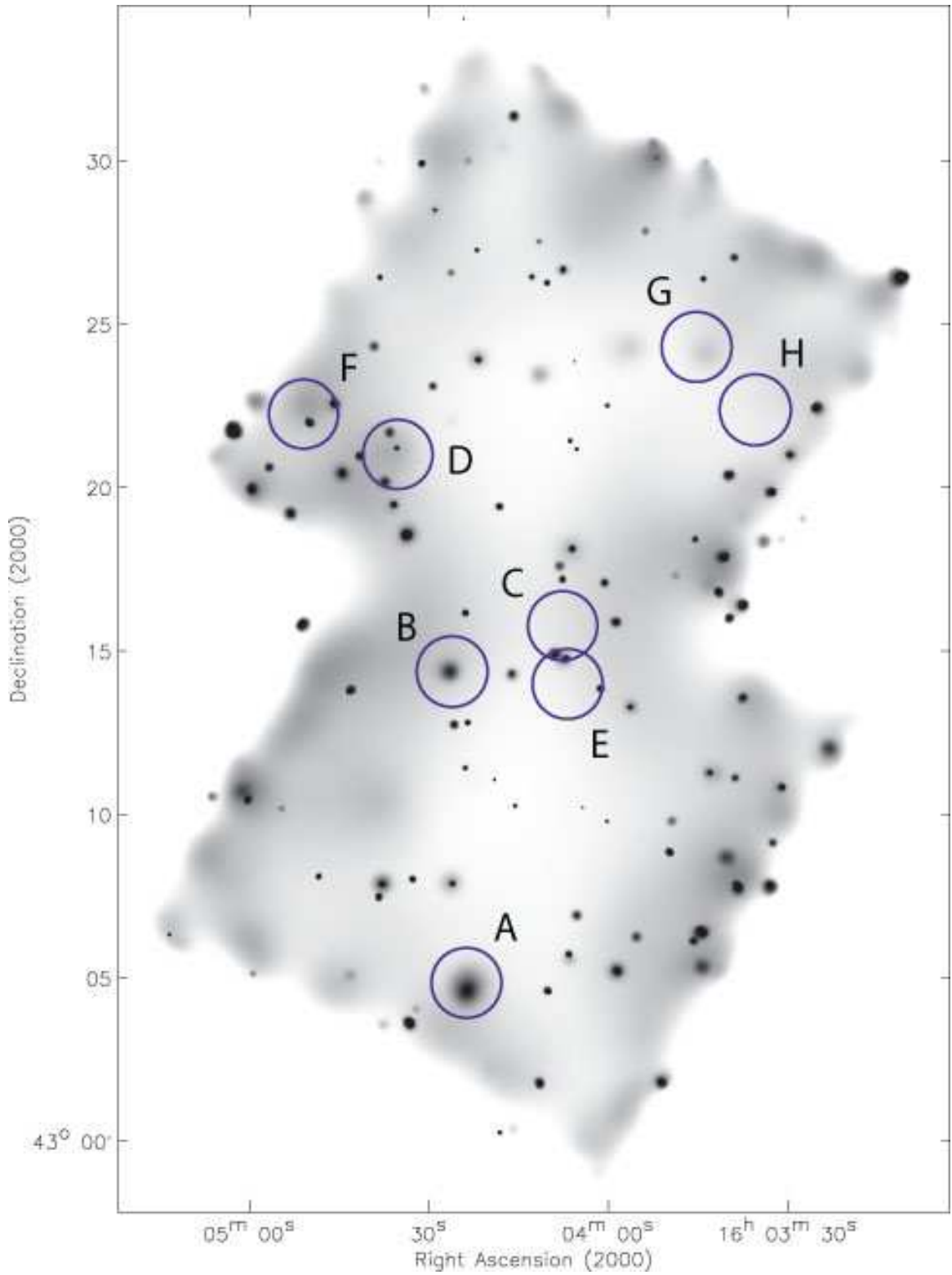}
\caption{Adaptively smoothed, exposure corrected mosaic of the Cl1604
  supercluster in the soft band (0.5-2 keV).  All sources are significant to
  at least the $3\sigma$ level. The location of eight red-galaxy
  overdensities, seven of which are spectroscopically-confirmed galaxy
  clusters or groups in the supercluster, are circled and labeled following the naming
  convention of Gal et al.~(2008) and that of Table \ref{tab-xprop}.  The
  circles have a radius of 0.5 $h_{70}^{-1}$ Mpc at the cluster redshifts. \label{fig-mosaic_asmoothed}}
\end{figure*}

All three datasets were reprocessed and analyzed using standard CIAO 3.3
software tools and version 3.2.2 of the {\it Chandra} calibration database
available through {\it Chandra} X-ray Center
(CXC)\footnote{http://cxc.harvard.edu/}.  New level 1 event files were
produced using the {\tt acis\_process\_events} script, which makes use of the
latest gain files and corrects for the effects of time-dependent variations
and charge transfer inefficiencies (CTI) in the ACIS CCDs.  Level 2 event
files were produced by filtering on standard ASCA grades (grades=0,2,3,4,6),
good status bits (status=0), and Good Time Intervals (GTIs) supplied by the
pipeline.  We checked the relative astrometry between the three event files
using the positions of 14 high signal-to-noise sources in the overlap region
between the two pointings and found the astrometric errors to be negligible.
Images for use in object detection and the examination of extended cluster
emission were created from the event lists in the 0.5-2 keV (soft), 2-8 keV
(hard) and 0.5-8 keV (full) X-ray bands with the standard 0.492
$^{\prime\prime}/$pixel binning.  To account for vignetting, CCD gaps, and
telescope dither effects we created spectrally-weighted, energy dependent
exposure maps in each of the three bands assuming a power-law source spectrum
with a photon index set to the slope of the X-ray background in the 0.5-8 keV
band, $\gamma = 1.4$ (Tozzi et al.~2001, Kushino et al.~2002).   

An adaptively smoothed, exposure corrected mosaic of the Cl1604 supercluster
in the soft band is shown in Figure \ref{fig-mosaic_asmoothed}.  The mosaic
was constructed by reprojecting each event file to the tangent point
$\alpha_{2000} = 16^{\rm h}04^{\rm m}12.4^{\rm s}$, $\delta_{2000} =
+43^{\circ}16^{\prime}18^{\prime\prime}$ with the {\tt reproject\_events}
task in CIAO and combining the resulting files.  Exposure variations and
vignetting were corrected for using a composite exposure map constructed by
reprojecting and combining our individual exposure maps with the {\tt
  reproject\_image\_grid}  task.   

It should be noted that since the aimpoints and roll-angles for the 7343 and
6933 observations are nearly identical, we have reprojected and combined the
images and exposure maps of these pointings and treat the combination as a
single observation throughout our analysis.  Hereafter we refer to the
composite pointing as Cl1604-North, while the 6932 observation will be
referred to as Cl1604-South.  Also, in what follows we largely work
separately with the Cl1604-North and Cl1604-South datasets as opposed to the
composite mosaic.

\subsection{Optical Imaging}

Our optical data of the system consists of two pointings of the
Large Format Camera (LFC) on the Palomar 5-m telescope and 17 pointings of
the higher-resolution Advanced Camera for Surveys (ACS) on the \emph{Hubble
  Space Telescope (HST)}.  Details of these observations and subsequent data
reduction are presented in Gal et al.~(2005, 2008) and Kocevski et al.~(2008) and are only briefly discussed here.

\begin{figure}[t]
\epsscale{1.0}
\plotone{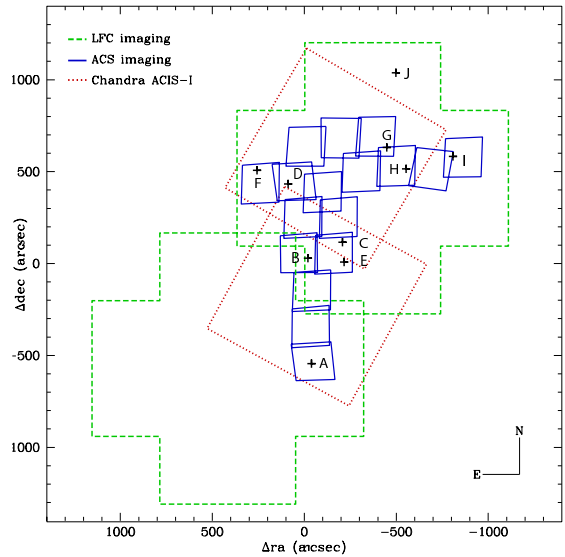}
\caption{Positions and FOVs of thpl two LFC, two ACIS-I, and 17 ACS pointings
  in the Cl1604 region. The positions of the ten red-galaxy overdensities found by Gal et
  al. (2005, 2008) are marked and labeled following the naming convention of Table
  \ref{tab-xprop}. \label{fig-data_outline}}
\end{figure}

The LFC is a mosaic camera of six $2048\times4096$ CCDs with a pixel scale of
0.182 $^{\prime\prime}$/pixel, resulting in an unvignetted FOV that is roughly circular with a 24$^{\prime}$ diameter.  We imaged two
pointings in the field of Cl1604 using the Sloan Digital Sky Survey (SDSS)
$r^{\prime}$, $i^{\prime}$, and $z^{\prime}$ filters, reaching a depth of
24.4, 24.2, and 23.2 mag in each band, respectively.  The area imaged by the
two LFC pointings relative to our ACIS-I imaging is shown in Figure
\ref{fig-data_outline}.  The ACS camera consists of two $2048\times4096$ CCDs
with a pixel scale of 0.05 $^{\prime\prime}$/pixel, resulting in a
$\sim3^{\prime}\times3^{\prime}$ FOV.  Our ACS imaging is comprised of a 17
pointing mosaic designed to image nine of the ten galaxy density peaks
observed in our LFC imaging of the supercluster.  An outline of the mosaic is
shown in Figure \ref{fig-data_outline}.  Observations were taken in both the
F606W and F814W bands, resulting in completeness depths of $\sim26.5$ mag in
each band.  The astrometry of the 17 pointings were fixed to that of the
USNO-B catalog (Monet et al.~2002) so as to match our LFC imaging and the
final images were resampled to a pixel scale of 0.03$^{\prime\prime}$/pixel.  Source catalogs were produced from both the LFC and
ACS imaging in each of the five bands observed using the {\tt SExtractor}
routine (Bertin \& Arnouts 1996) and cross-correlated to produce a single, composite optical catalog.
An outline of the region covered by our optical imaging relative the ACIS-I
observations is shown in Figure \ref{fig-data_outline}.

\subsection{Optical Spectroscopy}
\label{opt-spect}
The Cl1604 supercluster has been extensively mapped using the
Low-Resolution Imaging Spectrograph (LRIS; Oke et al. 1995) and the
Deep Imaging Multi-object Spectrograph (DEIMOS; Faber et al.~2003) on
the Keck 10-m telescopes (Oke, Postman, \& Lubin 1998; Postman, Lubin, \& Oke
1998; Lubin et al.~1998; Gal \& Lubin 2004; Gal et al.~2008). The
complex target selection, spectral reduction, and redshift measurements
are described in detail in Section 3 of Gal et al.~(2008). The final
spectroscopic catalog contains 1,618 unique objects. Redshifts derived for
these objects are given a spectroscopic quality, $Q_{spect}$, between 1 and
4, where 1 indicates a secure redshift could not be determined due to poor
signal, lack of features or reduction artifacts,  2 is a redshift obtained from either a single
feature or two marginally detected features, 3 is a redshift derived from at
least one secure and one marginal feature, and 4 is assigned to spectra with
redshifts obtained from several high signal-to-noise features.

In this sample, we find 202 stars and 1,381 extragalactic objects with redshifts of $z > 0.1$.  
A total of 449 galaxies are in the nominal redshift range of
the supercluster between $0.84 \le z \le 0.96$. This extensive
spectroscopic database is larger by a factor of $\sim 10$ than that
for any other known moderate-to-high redshift supercluster.

\section{Point Source Properties and Number Counts}
\label{sect-xps}

\subsection{Object Detection and Photometry}

We searched for point sources in the field of Cl1604 using the wavelet-based
{\tt wavdetect} procedure in CIAO.  We employed the
standard $\sqrt2$$^i$ series of wavelet pixel scales, with $i=0-16$.  These
scales are the radius of a Mexican Hat function in pixels with one pixel =
$0\farcs492$.  We also adopted a minimum exposure threshold of 20\% relative
to the exposure at the aimpoint of the observation and a threshold
significance for spurious detections of $10^{-6}$.  The latter implies less
than one false detection per ACIS chip (which contain $1024\times1024$ pixels
at full resolution), or roughly eight false detections over our entire field
of view (Freeman et al.~2002).  Object detection was carried out on the
unvignetting-corrected, full resolution images of the northern and southern
pointings seperately and in each of the soft, hard and full X-ray bands.  The
positions of the detected sources were then cross-correlated to produce a
multi-band source list for each pointing.  We set the final position of each
source in this composite list to the measured position in the band within
which the source was detected with the greatest significance.  With no cut on
significance, we detect a total of 99, 49, and 93 point sources in the soft,
hard, and full bands, respectively, in the northern pointing and 117, 54, and
105 sources in the same bands for the southern pointing.  Our detections are
summarized in Table \ref{tab-detections}.  In total, 265 unique sources were
detected in the two pointings of the Cl1604 system, of which 161 had
detection significances greater than $3\sigma$ in at least one X-ray band.

\begin{center}
\footnotesize
\begin{deluxetable}{ccccc}
\tablewidth{0pt}
\tablecaption{Point Source Detections in the Cl1604 supercluster \label{tab-detections}}
\tablecolumns{5}
\tablehead{ \colhead{Cl1604}  &  \colhead{Soft$^{1}$}  & \colhead{Hard$^{2}$} & \colhead{Full$^{3}$}  & \colhead{Total Uniq.}  \\   \colhead{Pointing}  &  \colhead{All ($3\sigma$)}  & \colhead{All ($3\sigma$)} & \colhead{All ($3\sigma$)}  & \colhead{All ($3\sigma$$^{4}$)}}
\startdata  
 North             &  99 (66)  & 49 (44) &  93 (75) &  123 (85) \\
 South             & 117 (67)  & 54 (42) & 105 (75) &  147 (90)
 \vspace*{-0.075in} \\ \\
\hline \vspace*{-0.075in} \\
 North+South       & 203 (120) & 94 (77) & 184 (136)  &   256 (161) \\
\vspace*{-0.075in}
\enddata                                                                                                                                       
\tablecomments{$^{1}$ 0.5-2 keV; $^{2}$ 2-8 keV; $^{3}$ 0.5-8 keV; $^{4}$ $3\sigma$ detection in at least one band}
\normalsize
\end{deluxetable}                                                                  
\end{center}                                                                        
\vspace*{-0.3in} 

Properties of the detected point sources, including count rates and detection
significances, were determined with follow-up aperture photometry on all
sources found by {\tt wavdetect}.  The apertures used were defined so as to
contain 95\% of the flux from a given point source.  As the {\it Chandra} PSF
is dependent on both energy and off-axis angle, we determined the 95\%
enclosed energy radius, $R_{95}$, at the position of each source using the
PSF libraries in the {\it Chandra} calibration database.  This was done at
1.49 and 4.51 keV for photometry in the soft and hard bands, respectively.
The radius of the resulting apertures at 1.49 keV ranged from $1\farcs8$
on-axis to $15\farcs8$ at $12\farcm1$ off-axis, the distance of our most
off-axis source; the latter increased to $17\farcs2$  at 4.51 keV.  The
background near each source was determined in an annulus extending from 1.2
to 2.4 $\times$ $R_{95}$, with appropriate masking of nearby sources when
necessary.  We carried out the photometry on the vignetting-corrected, soft- and
hard-band images of both pointings and an aperture correction of $1/0.95$ was
applied to the background-subtracted net counts of each source.  Counts in
the full band were then determined as a sum of the measured net counts in the soft and hard bands.  

\begin{figure}
\epsscale{1.0}
\plotone{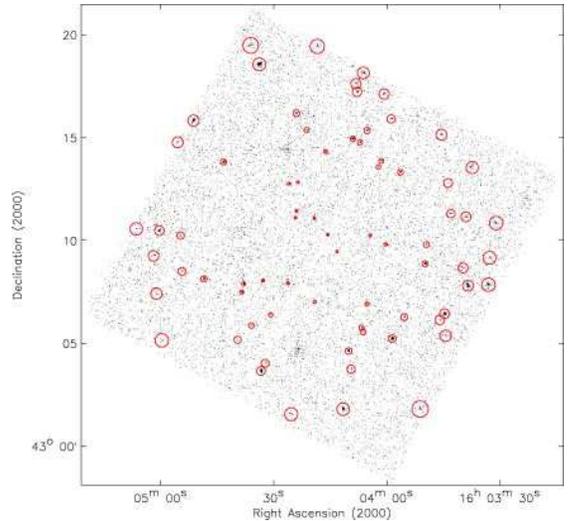}
\caption{Apertures containing 95\% of the flux from point sources detected in
  the Cl1604-South pointing with a $3\sigma$ significance in the 0.5-2 keV
  band.  The radius has been calculated at an energy of 1.49 keV.  The
  underlying soft-band image has been binned to a pixel scale of 2
  $^{\prime\prime}$/pixel. \label{fig-aps}}
\end{figure}

As the low number of counts for many of the detected sources are not
favorable to a full spectroscopic analysis, we determined the soft- and
hard-band fluxes of each source by normalizing a power-law spectral model to the
net count rate measured for each source.  These rates were determined by
dividing the net counts measured in the vignetting-corrected images by the
nominal exposure time at the aimpoint of each observation.  We assumed a
photon index of $\gamma = 1.4$ for the power-law model and a Galactic neutral
hydrogen column density of $1.21\times10^{20}$ cm$^{-2}$ (Dickey \& Lockman
1990), resulting in a count rate to unabsorbed-flux conversion factor of
$5.85\times10^{-12}$ erg cm$^{-2}$ cnts$^{-1}$ and $2.08\times10^{-11}$ erg
cm$^{-2}$ cnts$^{-1}$ in the soft and hard bands, respectively.  Full-band
fluxes were again determined by summing the flux in the soft and hard bands.  

\begin{figure*}
\epsscale{1.1}
\plotone{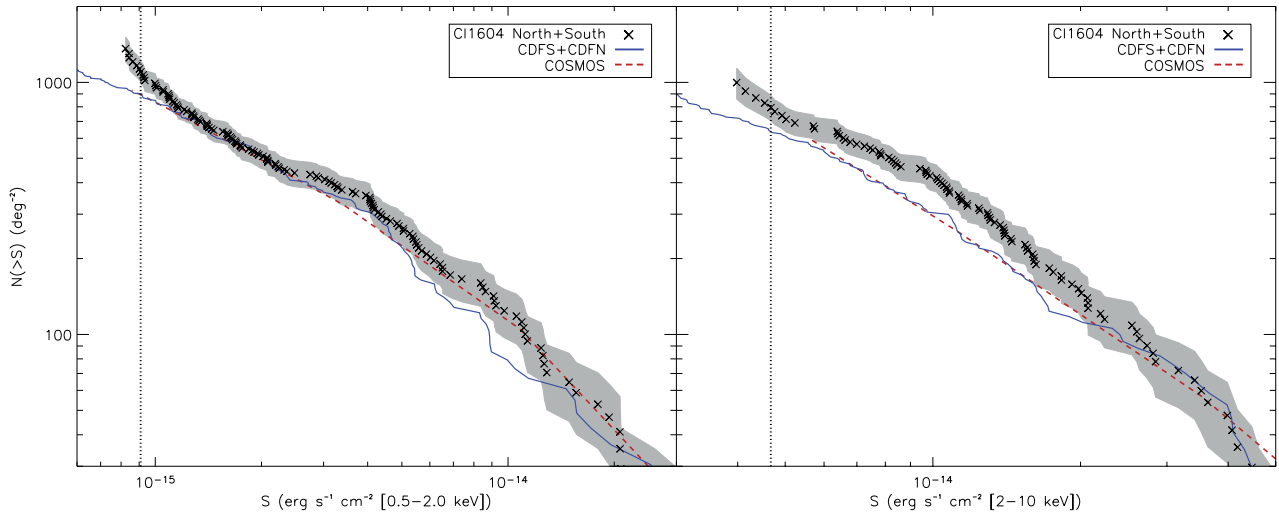}
\caption{Combined cumulative source number counts versus flux for the
  northern and southern pointings of the Cl1604 supercluster in the soft (0.5-2 keV;
  \emph{left}) and hard band (2-10 keV; \emph{right}).  The shaded region denotes a
  $1\sigma$ variance in the number counts.  Only sources detected above the
  $3\sigma$ level are included.  The results of an \emph{XMM-Newton} survey of the COSMOS
  field (Cappelluti et al.~2007) and that of a 130 ksec \emph{Chandra} observation
  of the CDFS and CDFN are shown for comparison (dashed and solid line,
  respectively). The vertical dotted line represents the flux at which our
  sky coverage dropped to 20\% of the full ACIS-I FOV. \label{fig-lognlogs}}
\end{figure*}

Properties for the 161 unique sources detected with at least a $3\sigma$
significance in the field of Cl1604 are listed in Table \ref{tab-sources}.
The table includes source IDs (Column 1), right ascensions and declinations
(Columns 2-3), positional errors determined using the empirical relationships
of Kim et al.~(2007; Column 4), aperture-corrected, net counts above the
background in the soft, hard and full bands (Columns 5-7), X-ray fluxes in
all three bands in units of erg s$^{-1}$ cm$^{-2}$ (Columns 8-10), detection
significances measured as 

\begin{equation}
\label{eqsig}
\footnotesize
   sig = net\_cnts / \left(1.0 + \sqrt{0.75 + bkg\_cnts}\right)
\normalsize
\end{equation}

\noindent (Column 11-13), hardness ratios measured as $HR = (H-S)/(H+S)$
where $H$ and $S$ are the net counts in the hard and soft bands,
respectively (Column 14), and a three letter string indicating the bands
in which the source was detected, with S, H and F indicating the soft, hard and full bands, respectively (Column 15).

\subsection{Log N - Log S}

To investigate whether an excess of unresolved X-ray sources exist in the
field of Cl1604 relative to fields without such a structure we have
calculated the cumulative source number counts, $N(>S)$, as described by Gioia et al.~(1990) using

\begin{equation}
   N(>S) = \sum_{i=1}^{N} \frac{1}{\Omega_{i}} {\rm deg}^{-2} 
\end{equation}

\noindent Here $N$ is the total number of detected point sources and
$\Omega_{i}$ is the sky area in square degrees sampled by the detector down
to the flux of the $i$th source.  The variance of the number counts was in turn calculated as

\begin{equation}
   \sigma_{i}^{2} = \sum_{i=1}^{N} \left(\frac{1}{\Omega_{i}}\right)^{2}
\end{equation}

Determining $\Omega_{i}$ for a given \emph{Chandra} observation is
complicated by the fact that the flux limit across the ACIS-I array varies
due to vignetting and PSF degradation as a function of off-axis angle.  In
order to calculate $\Omega_{i}$ we constructed a flux limit map using a
similar method to that employed by Johnson et al.~(2003) and Cappelluti et
al. (2005).  First, all point sources detected by {\tt wavdetect} (at all
significance levels) in our two pointings were replaced with an estimate of
the local background with the CIAO tool {\tt dmfilth}.  These images were
then binned to a pixel scale of $32^{\prime\prime}$/pixel to produce a coarse
background map.  Given Equation \ref{eqsig}, the flux limit, $S_{lim}$, for a
$3\sigma$ point source detection in any one pixel is then

\begin{equation}
   S_{\rm lim} = \frac{3 V k}{t} (1.0+\sqrt{0.75 + B \pi R_{95}^{2} / A_{\rm pix}}),
\end{equation}

\noindent where $B$ is the background level in counts, $A_{\rm pix}$ is the
sky area covered by one binned pixel, $R_{95}$ is the aforementioned 95\% enclosed
energy radius used for our aperture photometry, $V$ is a
vignetting correction factor, $k$ is the count rate to flux conversion factor,
and $t$ is the exposure time at the aimpoint of the observation.  The values
for $R_{95}$ are determined for each pixel given its off-axis angle, while
$V$ is estimated by normalizing an exposure map to its maximum value at the
aimpoint of the observation.  Using this flux limit map we can calculate
$\Omega_{i}$ by summing the sky area covered by all pixels with $S_{\rm lim}$
equal to or greater than the flux of the $i$th source.   An important point
to note is that we have not removed or masked any diffuse cluster emission
and have instead treated it as an enhanced local background in constructing
the flux limit map.  This allows us to properly account for the fact that our
flux limit for point source detection is effectively increased in the
presence of diffuse emission when we calculate the $1/\Omega_{i}$ corrective weights.

The combined cumulative source number counts for the northern and
southern pointings of the Cl1604 supercluster are shown in Figure \ref{fig-lognlogs}.  
The distribution is shown in the 0.5-2 keV (left panel) and 2-10 keV (right panel) bands.
The latter was chosen to ease comparison with previous studies and was
obtained by extrapolating our 2-8 keV fluxes to 10 keV; we refer to 2-10 keV band as the hard$_{10}$ band hereafter.  Also shown are the
cumulative number counts measured in the COSMOS field (Scoville et al.~2007) and the combined counts
obtained in the {\it Chandra} Deep Field South and North (CDFS and CDFN;
Rosati et al.~2002; Brandt et al.~2002).  The COSMOS results are those of
Cappelluti et al.~(2007) converted to a spectral index of $\gamma = 1.4$, while
the combined CDFS and CDFN counts are the result of our own reanalysis of two
single pointings in each field. Data for these fields were obtained from the
{\it Chandra} archive\footnote{http://cxc.harvard.edu/} and analyzed in an
identical manner to that of the Cl1604 field.  The two pointings used have
observation ID numbers of 581 and 2232; each is a single pointing of the
ACIS-I array with exposure times of roughly 130 ksec. We have combined the
number counts obtained for each field separately into a unified reference
field (hereafter CDF) to which we compare our source counts throughout.

\begin{figure}
\epsscale{1.0}
\plotone{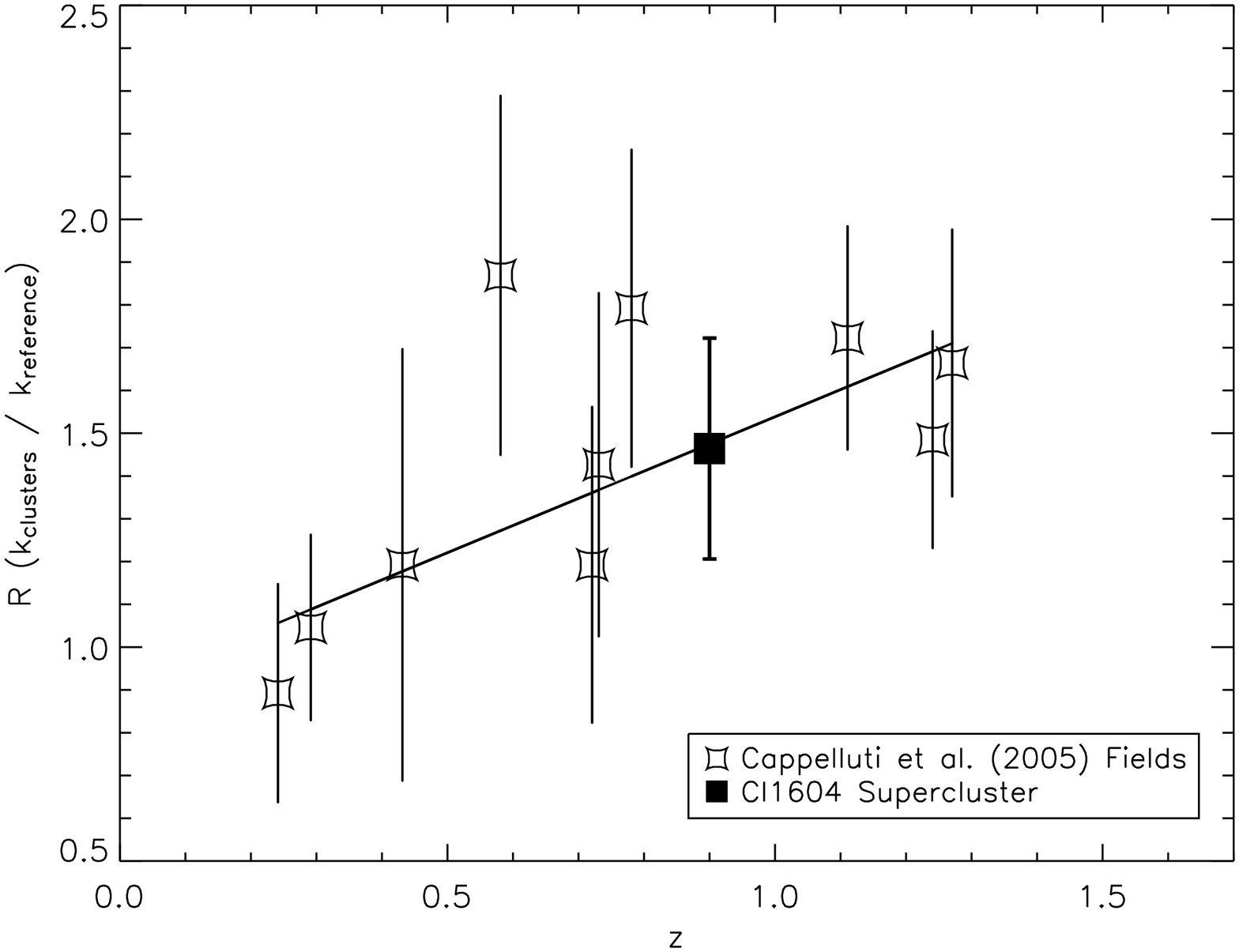}
\caption{Overdensity amplitude as a function of cluster redshift observed by
  Cappelluti et al.~(2005) in the field of 10 clusters in the hard$_{10}$ band.
  Overplotted is the overdensity we measure in the field of the Cl1604
  supercluster.  We find that the Cl1604 counts are in excellent agreement
  with their best-fit linear relation. (Adapted from Figure 6 of Cappelluti et al.~2005) \label{fig-capp}}
\end{figure}

In the soft band we find that the number of sources detected in the Cl1604
pointing and the composite blank field are statistically consistent ($<1\sigma$
difference) over a majority of the sampled flux range.  The greatest
deviation of the Cl1604 counts is observed at $9.1\times10^{-15}$ erg s$^{-1}$ cm$^{-2}$, where we find a
$2.0\sigma$ excess in the supercluster field relative to the CDF counts.  This is likely due to the
underdensity of soft sources previously reported at these fluxes in the CDFS (Yang et al.~2003)
which suppresses the combined CDF number counts at this flux.  When our
measured source density is compared to that of the COSMOS field, which covers a
significantly larger area than the CDFS, we find
excellent agreement at the bright end of the distribution.  We also observe a deviation from
both the CDF and COSMOS fields near our $3\sigma$ flux limit, but we caution that at these fluxes
the source counts are heavily weighted due to a rapidly decreasing sky coverage.
At a flux of $9.1\times10^{-16}$ erg s$^{-1}$ cm$^{-2}$ the effective detector
area drops to 20\% of the total ACIS-I FOV.  As is common in the literature,
we adopt this flux as our effective minimum for this analysis; the flux is
denoted by a vertical dotted line in Figure \ref{fig-lognlogs}.  

Unlike the general agreement found between the supercluster and blank field counts in the soft band, we find that the source density 
in the hard$_{10}$ band significantly surpasses the density measured in both the reference and COSMOS fields. 
As a result of a steeper power-law slope, the Cl1604 counts deviate from the
blank field distribution near $2\times10^{-14}$ and reach a maximum excess at
$9.7\times10^{-15}$ erg s$^{-1}$ cm$^{-2}$, where we find a $2.5\sigma$
overdensity relative to the CDF counts.  Below this flux a break in the
source counts is observed and the slope of the distribution flattens,
reducing the difference between the supercluster and blank field counts.  At
our 20\% FOV flux limit of $4.7\times10^{-15}$ erg s$^{-1}$ cm$^{-2}$ the excess is reduced to
a $1.7\sigma$ deviation.

To parameterize our cumulative flux distribution, we have fit the unbinned
soft- and hard$_{10}$-band counts in the Cl1604 field with a power-law model of the form 

\begin{equation}
\label{powlaw}
\footnotesize
   N(<S) = k \left(\frac{S}{S_{0}}\right)^{-\alpha} ({\rm deg}^{-2})
\normalsize
\end{equation}

\noindent using the maximum likelihood method described by Crawford et al.~(1970) and
Murdoch et al.~(1973).  The normalization, $k$, was determined by requiring
that the best fit slope, $\alpha$, reproduce the number of sources observed
at the flux limit.  Our best-fit single power-law model in the soft band is 

\begin{eqnarray}
\label{powlaw_results_soft}
\footnotesize
   N_{0.5-2}(<S) = 550 \left(\frac{S}{2\times10^{-15}}\right)^{-0.88\pm0.08}  \nonumber
\normalsize
\end{eqnarray}
   
In the hard$_{10}$ band there is a clear break at $9.2\times10^{-15}$ erg s$^{-1}$
cm$^{-2}$, below which the slope of the distribution flattens, in qualitative
agreement with previous deep Chandra and XMM surveys (Brandt \& Hasinger 2005). 
Performing a maximum likelihood fit to the bright-end counts above the break,
we obtain a power-law model of

\begin{eqnarray}
\label{powlaw_results_hard}
\footnotesize
N_{2-10}(<S) =   419 \left(\frac{S}{1\times10^{-14}}\right)^{-1.59\pm0.21}.  \nonumber
\normalsize
\end{eqnarray}

\noindent Our best-fit slope of $\alpha=1.59\pm0.21$ over the flux range of $1\times 10^{-14}$ to
$4\times 10^{-14}$ erg s$^{-1}$ cm$^{-2}$ is significantly steeper than the
slopes found over similar fluxes in the CDFN ($1.0\pm0.3$; Brandt et
al. 2002) and the CDFS ($0.61\pm0.1$; Rosati et al.~2001).  This difference in
the slope of the cumulative distribution is responsible for building up the
$2.5\sigma$ source excess observed at $\sim1\times 10^{-14}$ erg s$^{-1}$ cm$^{-2}$.

It appears that over the FOV of nearly two full ACIS-I pointings covering the
Cl1604 supercluster we find no significant excess of soft X-ray sources
relative to a blank field, but a clear overdensity of harder, presumably
more obscured sources is detected.   Cappelluti et al.~(2005) previously reported a
correlation between the amplitude of the overdensity observed in the
hard$_{10}$ band and cluster redshift.  Normalizing our observed source density to the mean amplitude of their reference field counts at
$1\times10^{-14}$ erg s$^{-1}$ cm$^{-2}$, we find the Cl1604 field exhibits
a factor 1.47 overdensity of hard$_{10}$-band sources.  In Figure \ref{fig-capp} we plot
the overdensity amplitude as a function of cluster redshift observed by
Cappelluti et al.~(2005) along with the overdensity amplitude in the Cl1604
field.  We find that the Cl1604 counts are in excellent agreement with their best-fit linear relation.

It has also been noted that the observed field-to-field variation in the hard band
is substantially above the level expected due to Poisson noise (Cowie et
al. 2002), suggesting the variations are due to intrinsic clustering of the
population and their tracing of underlying large-scale structures.  If the
excess sources we have detected are associated with the Cl1604 supercluster
then this observation, along with the fact that we do not see such an
overdensity in the soft band, seems to support the notion that more obscured,
harder X-ray sources are more highly biased tracers of LSS than their softer
counterparts.  This issue has been debated with Yang et al.~(2003) and
Basilakos et al.~(2004) finding an increased angular clustering amplitude of
hard-band sources while Gilli et al.~(2005) and Yang et al.~(2006) find no
dependence of the clustering scale length with X-ray spectral shape (see also Miyaji et al.~2007).

Finally, it is important to stress two points regarding the nature of the
observed overdensity.  First, if the excess sources are at the supercluster
redshift of $z=0.9$ , then their luminosities strongly suggest the sources
are AGN as opposed to powerful starburst galaxies.   For example, a source
detected at our $3\sigma$ flux limit would have a rest-frame 0.5-8 keV
luminosity of $8.5\times10^{42}$ $h_{70}^{-2}$ erg s$^{-1}$, far above the
luminosity attributable to starburst galaxies (e.g. Bauer et al.~2002),
making it highly unlikely that the source overdensity is due to increased
starburst activity in the supercluster.  Secondly, we point out that the
observed overdensity persists over nearly two full ACIS-I pointings.
Previously reported source overdensities have often been presented on a
chip-by-chip basis ($8^{\prime}\times8^{\prime}$; e.g. D'Elia et al.~2004,
Cappelluti et al. 2005).  On such scales Poisson noise may contribute a
significant signal to fluctuations observed in the source counts (Cappelluti
el al. 2007).  In fact Cappelluti et al.~(2005) note that their observed
excesses disappear when their source counts are integrated over an entire
ACIS-I FOV.  This is in agreement with Kim et al.~(2004b) who find no
statistically significant difference between cluster and cluster-free fields
on scales of $16^{\prime}\times16^{\prime}$.  The fact that our larger FOV
has not smoothed away our observed overdensity suggests the increased
amplitude is not due to a statistical fluctuation, but instead due to sources
which are indeed associated with the supercluster, whose constituent clusters
extend over the entire FOV of the two pointings.  In \S\ref{sect-z} we
bolster this claim with evidence for a peak in the redshift distribution of
X-ray sources near that of the Cl1604 system.

\subsection{Matching to Optical Sources}
\label{sect-opt-matching}

To determine the origin of the overdensity observed in the number counts and
to ascertain whether the sources are truly associated with the Cl1604
supercluster we searched for optical counterparts to all 161 unique point sources detected
in the field of Cl1604 above the $3\sigma$ level in at least one of the three
X-ray bands.  Matching of our X-ray source list to the LFC+ACS optical catalog was carried
out using the maximum likelihood technique described by Sutherland \&
Saunders (1992) and more recently implemented by Taylor (2005) and Gilmour et al.
(2007).  The method gauges the likelihood that a given optical object is
matched to an X-ray source by comparing the probability of finding a genuine
counterpart with the positional offset and magnitude of the optical candidate
relative to that of finding a similar object by chance.  The key statistic
used is the known as the likelihood ratio, ${\rm LR_{i,j}}$, which specifies
the likelihood that X-ray source $j$ is associated with optical object
$i$.  If we assume a Gaussian form for the probability distribution of the
X-ray positional errors, it can be defined as

\begin{equation}
\label{eq-LR}
   {\rm LR}_{i,j} = \frac{e^{-r^{2}_{i,j} / 2\sigma^{2}_{j}}}{\sigma^{2}_{j} n(< m_{i})},
\end{equation}

\noindent where $r_{i,j}$ is the positional offset between optical object $i$
and X-ray source $j$, $n(<m_{i})$ is the number density of optical objects
brighter than object $j$ in the F814W ($i^{\prime}$) band in the ACS (LFC)
catalog, and $\sigma_{j}$ is the positional error associated with X-ray
source $j$.   

The advantages of this approach over a simple nearest neighbor match is that
${\rm LR_{i,j}}$ takes into account the density of objects as faint as the
optical counterpart as well as the distance between sources and the X-ray
positional errors.   This is vitally important when matching to a source list
drawn from deep, high-resolution ACS imaging, which has an extremely high
surface density of faint sources.  Without consideration of magnitude an
inordinate number of X-ray sources would be matched to optical objects near
the detection limit of the catalog.  Instead, when surface density is taken
into account, brighter sources, which are rarer in the optical catalog, are
given an increased likelihood of being a genuine counterpart to an X-ray
source compared to fainter sources with the same positional offset.

We considered an optical source a candidate counterpart if it fell within the error circle of an X-ray source and calculated ${\rm {LR}_{i,j}}$ for each candidate-X-ray pair.  Since the \emph{Chandra} PSF degrades with increased distance from the aimpoint of the observation, the positional uncertainty associated with each X-ray source depends on its off-axis angle.  To account for this effect we have determined the errors on our X-ray positions using the empirically derived relations of Kim et al.~(2007), who find that centroiding errors increase exponentially with off-axis angle and decrease as the source counts increase with a power-law form.  To simplify the calculation of ${\rm {LR}_{i,j}}$ we have assumed that any optical positional errors are negligible compared to the larger X-ray uncertainties, which ranged from $0\farcs5$ on-axis to $12\farcs0$ at $12\farcm1$ off-axis for a source detected with 10 counts\footnote{The quoted errors are at the 95\% confidence level}.

To calculate $n(<m_{i})$ for a given optical counterpart, we determined the
magnitude distribution of sources in the same subset of the composite optical
catalog from which the candidate source was drawn.  If the source fell within
the boundaries of the ACS mosaic, $n(<m_{i})$ was determined from the density
and magnitude distribution of ACS detected objects.  A similar procedure was
used for sources only detected in the LFC imaging.  

While ${\rm LR_{i,j}}$ provides the likelihood that X-ray source $j$ is
associated with optical object $i$, the ratio itself does not provide an
estimate of the reliability of a given match.  This can be obtained by
comparing the value of ${\rm LR_{i,j}}$ for a particular match to the
distribution of ${\rm LR_{i,j}}$ values for chance matches.  We determined
the latter by randomizing the positions of each X-ray source 10000 times and
recording the distribution of ${\rm LR_{i,j}}$ values for the resulting
chance matches to optical sources.  Again, this process was done separately
for X-ray sources falling within and outside the ACS imaging.  Following
Gilmour et al.~(2007), the reliability of a match was defined as the probability of not obtaining ${\rm LR_{i,j}}$ randomly,

\begin{equation}
\label{eq-R}
   R_{i,j} = 1 - \frac{\Sigma {\rm N(LR_{j} > LR_{i,j})}}{10000},
\end{equation}

\noindent where ${\rm N(LR_{j} > LR_{i,j})}$ is the number of chance matches
in our Monte Carlo simulation with likelihood ratios that exceeded ${\rm
  LR_{i,j}}$.  Here $R_{i,j}$ is essentially the binomial probability that a
given match with a specific value of ${\rm LR_{i,j}}$ is a true association
and not a chance match.  

Given the relatively large X-ray positional uncertainties and the high source density in
our ACS imaging, there are often several candidate counterparts to any given
X-ray source and many of these can have high reliability estimates.  Following
Rutledge et al.~(2000) we can use $R_{i,j}$ to determine which optical
candidate is the genuine counterpart and the probability that there is
instead no genuine counterpart.  Since $R_{i,j}$ is the the probability that a given match is a
true association, the probability that optical source $i$ is the genuine
counterpart to X-ray source $j$ is

\begin{equation}
\label{eq-probk}
   P_{i,j} = \frac{R_{i,j} \Pi_{k\neq i}^{M} (1 - R_{k,j})}{S}.
\end{equation}

\noindent Here $M$ is the total number of optical candidates and $S$ is a
normalization factor defined below.   $P_{i,j}$ is simply the product of the probability that the $i$th
optical object is a true association ($R_{i,j}$) with the probabilities that the
remaining optical candidates are not ($1 - R_{k\neq i,j}$).  Likewise, the
probability that there is no genuine counterpart given $M$ candidates is

\begin{equation}
\label{eq-probnone}
   P_{none,j} = \frac{\Pi_{k=1}^{M} (1 - R_{k,j})}{S}
\end{equation}

\noindent where $S$ is a normalization factor that varies for each source to insure
that the probabilities $P_{i,j}$ and $P_{none,j}$ sum to unity.

Following Gilmour et al.~(2007), we consider an X-ray source with a single
candidate counterpart to be matched to a given optical object if $P_{i,j} >
0.8$ for objects in the ACS catalog (i.e. the probability for a genuine match
is four times that of a null match).  This cutoff was reduced to 0.75 for
matches with LFC objects to include several well aligned sources near the
completeness limit of the LFC catalog.  For sources with more than a single
candidate counterpart, the genuine optical match was defined as the source
with $\Sigma_{k} P_{k,j} > 0.8$ and $P_{i,j} / \Sigma_{k\ne i} P_{k,j} > 4$,
where $k$ is the set of optical candidates.  If no single candidate
fulfilled this requirement, yet their summed probabilities exceeded 0.8, we
considered all optical objects with $P_{i,j} > 0.2$ as possible counterparts.

Using this prescription we have matched 100 X-ray sources with unique optical
counterparts found in a our LFC and ACS imaging of Cl1604.  An
additional 11 sources were found to have two candidate counterparts, while
one source is matched to three optical candidates.  A remaining 49 X-ray
sources were found to have no likely optical counterpart within the limits of
our optical imaging. The X-ray and optical coordinates of the matched sources
are listed in Table \ref{tab-matches}.  The table includes source IDs (Column
1), X-ray centroids (Columns 2-3), optical centroids (Columns 4-5), the
probability the given X-ray source has an optical counterpart ($\Sigma_{k}
P_{k,j}$; Column 6), the probability that the given optical source is the
genuine counterpart ($P_{i,j}$; Column 7), redshift (Column 8), and the
$Q_{spect}$ value for the derived redshift (Column 9).

\subsection{Supercluster Members}
\label{sect-z}

As described in \S\ref{opt-spect}, our extensive spectroscopic database
contains redshifts for 1381 galaxies and spectra of 202 stars in the field of
the Cl1604 supercluster.  Of the 125 optical sources matched to our X-ray
catalog, a total of 42 have spectroscopic information available.   All but two
of these are sources considered genuine optical matches using the criteria
put forth in \S\ref{sect-opt-matching}, while the remaining pair are both one
of two candidate counterparts to their respective X-ray source.  We have
derived reliable redshifts with $Q_{spect} > 2$ for 35 of the 42 sources which have
spectroscopy.  While redshifts for all 42 sources are listed in Table \ref{tab-matches}, we
only make use of the 35 $Q_{spect} > 2$ redshifts for the following discussion.

\begin{figure}
\epsscale{1.0}
\plotone{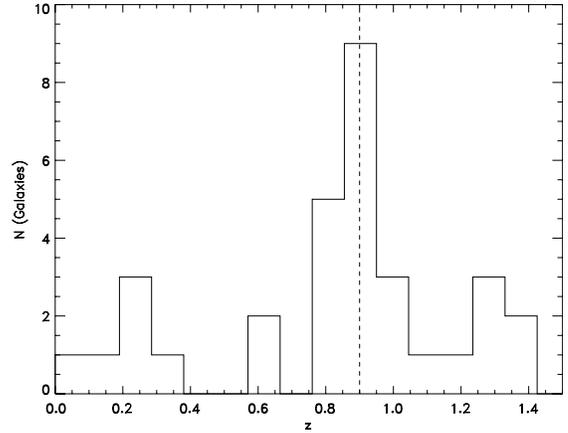}
\caption{Redshift distribution of the 34 galaxies matched to X-ray point
  sources in the field of the Cl1604 Supercluster.  Two source at $z=2.04,
  2.30$ fall outside the plot range.  We observe a peak in the redshift
  distribution centered on the mean redshift of the Cl1604 system, which is
  denoted by the vertical dashed line.\label{fig-zhist}}
\end{figure}

We find that one of the X-ray sources has a clear stellar spectrum, while the
remaining 34 sources are extragalactic covering the redshift range $0.055 \le
z \le 2.30$.  The redshift distribution of the extragalactic sources is
shown in Figure \ref{fig-zhist}.  The distribution exhibits a clear peak at z=0.9, the
average redshift of the Cl1604 supercluster.  A total of 9 sources have redshifts between
$0.84<z<0.96$ and fall within the traditional redshift boundaries of the supercluster
as defined by Gal et al.~(2004).  An additional 6 are found in the immediate
foreground of the system between $0.80<z<0.84$ and another 2 are found
immediately behind the supercluster at $0.96<z<1.0$.  The remaining sources
are high redshift (10 at $z>1.0$) or foreground sources (9 at
$z<0.8$) seen in projection.   

We find that 4 of the 9 supercluster members have hard$_{10}$-band fluxes
between 1 and 2 $\times10^{-14}$ erg s$^{-1}$, the flux range where the
source excess is observed.  The same is true for 2 of the 6 sources in the
immediate background and foreground of the supercluster, 4 of the 9 sources
at $z<0.8$ and 3 of the 10 sources at $z>1.0$.  Combined, these sources
account for 31\% of the sources that contribute to the overdensity, leaving
an additional 29 sources with hard$_{10}$-band fluxes between 1 and 2
$\times10^{-14}$ erg s$^{-1}$ for which we do not yet have measured redshifts.

\begin{figure}
\epsscale{1.15}
\plotone{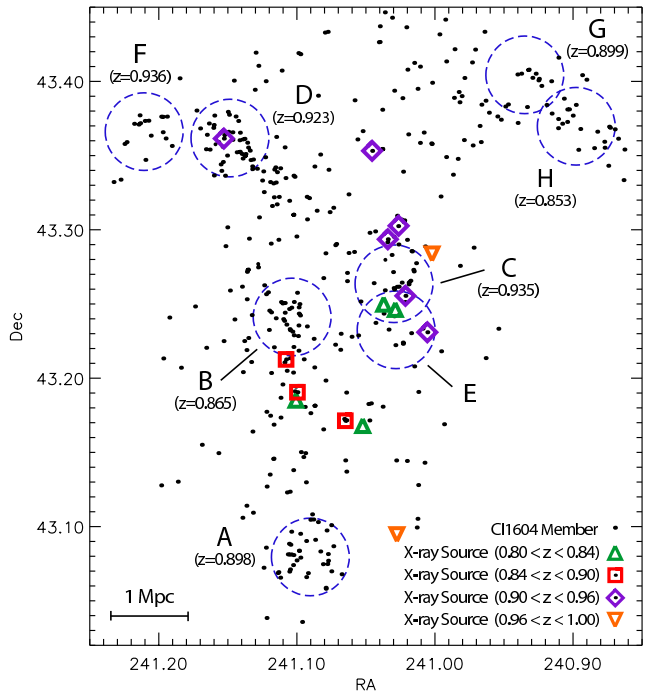}
\caption{Spatial distribution of the 9 X-ray sources that fall within the
  traditional redshift boundaries of the Cl1604 supercluster
  ($0.84<z<0.96$). We have split the sample into two redshift bins in order
  to highlight cluster membership.  Also shown are the locations of the 4 and
  2 sources in the immediate foreground ($0.80<z<0.84$) and background
  ($0.96<z<1.0$) of the system, respectively.  The cluster labels follow the
  naming convention of Table \ref{tab-xprop}. See text for
  details. \label{fig-spatdist}}
\end{figure}

In Figure \ref{fig-spatdist} we plot the spatial distribution of the 9
supercluster members and the additional 6 sources that fall in the immediate
foreground and background of the supercluster which may be associated to the
large-scale structure of the system.  Also shown are the locations and
systemic redshifts of individual clusters in the system.  We find 3 supercluster members
associated with Cl1604+4314 (hereafter Cluster B) at $z=0.865$; the two
sources closest to the cluster center have nearly identical redshifts to that of the cluster ($z=0.867,0.871)$, while the
one with the greatest projected distance is at a higher redshift ($z=0.899$).  The
two sources found in the foreground of Cluster B are themselves clustered in redshift
space at $z=0.828, 0.824$.  Near Cl1604+4316 (Cluster C) at $z=0.935$
we find 4 nearby sources that are associated in redshift space having
$z=0.927, 0.935, 0.937, 0.935$.  An additional source is located in the
foreground of the cluster at $z=0.913$.  In Cl1604+4321 (Cluster D) we find one
source with a redshift identical to that of the systemic cluster redshift of
$z=0.923$.  

Despite extensive spectroscopic coverage of Cl1604+4304 (Cluster A) we find
no sources associated with the system. The cluster is the richest system in
the Cl1604 association and it is also the most X-ray luminous.  We also do not find any sources near the center of
Cluster B, with the nearest source roughly 1 $h_{70}^{-1}$ Mpc from the
cluster center.  As we discuss in \S\ref{sect-diff}, these systems are the
only clusters for which we detect diffuse emission.  Although this emission
effectively raises our point source detection threshold in the soft band, the emission is
relatively low-level and would only mask the faintest of sources.  For example, in the center of Cluster A we find that the
flux limit for a $3\sigma$ detection rises to $1.13\times10^{-15}$ erg
s$^{-1}$ in the soft band.  Therefore, the diffuse emission from these systems does not
prevent us from detecting bright point sources in the soft band, nor does it
effect our ability to detect sources in the hard band, yet none are found,
either with \emph{wavdetect} or by visual inspection.  This is quite unlike
Clusters C and D, in which we find sources near the center of each system
(both in projection and redshift space).  This observation is consistent with
the results of Gilmour et al.~(2007) who found AGN avoided the densest
regions of the low redshift supercluster Abell 901/902.  

It is interesting to note that Gal et al.~(2008) found that Clusters B and D
showed evidence for velocity segregation in the redshift distribution of
their member galaxies, indicating they may have undergone a recent merger
event or have a significant population of actively accreting galaxies.  This
is quite unlike Cluster A, which is the most relaxed system in the
supercluster. The difference in the dynamical activity of Clusters A and
B indicate the global properties of these systems may affect the level of AGN activity observed in each
cluster.  In a forthcoming paper we will discuss the environments and optical properties of the supercluster
members in greater detail, including the local galaxy density near the host
galaxies and the global properties of the clusters within which
they reside.  This future study will further explore the possibility that
cluster properties, such as increased dynamical activity or the presence of a
significant intracluster medium (ICM), can affect the level of AGN activity
observed in a system.

\section{Diffuse X-ray Emission}
\label{sect-diff}

\begin{figure*}
\epsscale{1.0}
\plotone{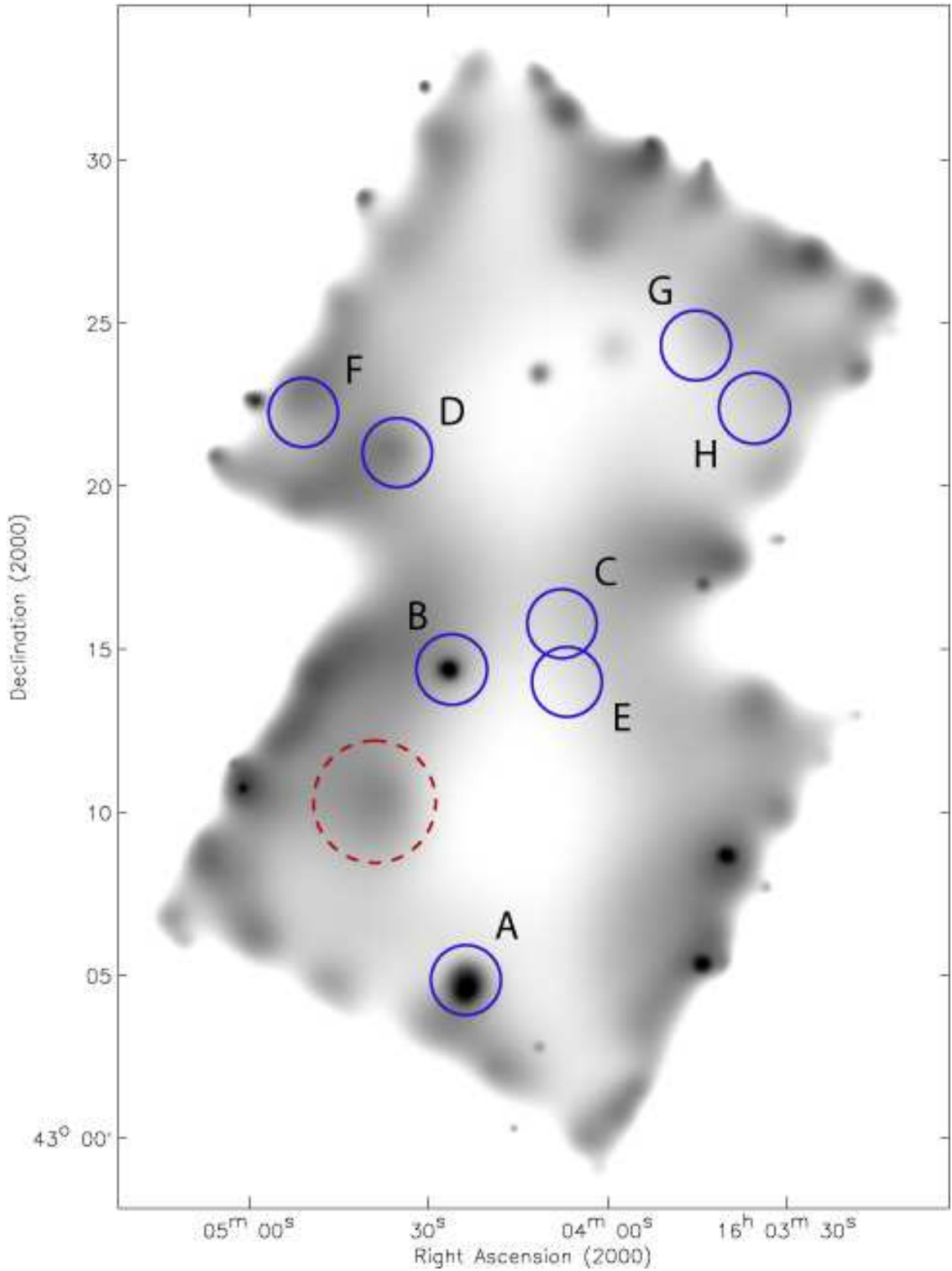}
\caption{Extended emission in the field of the Cl1604 supercluster in the
  0.5-2.0 keV band.   All point sources detected with {\tt wavdetect} were
  removed and the image adaptively smoothed with a minimal smoothing scale of
  $10^{\prime\prime}$.  The location of eight red-galaxy
  overdensities, seven of which are spectroscopically-confirmed galaxy
  clusters or groups in the supercluster, are circled and labeled following the naming
  convention of Gal et al.~(2008) and that of Table \ref{tab-xprop}.  The
  circles have a radius of 0.5 $h_{70}^{-1}$ Mpc at the cluster redshifts.  A
  serendipitously detected foreground cluster at $z=0.3$ is noted by the
  dashed circle. \label{fig-mosaic_asmoothed_nops}}
\end{figure*}

\begin{figure*}
\epsscale{1.0}
\plotone{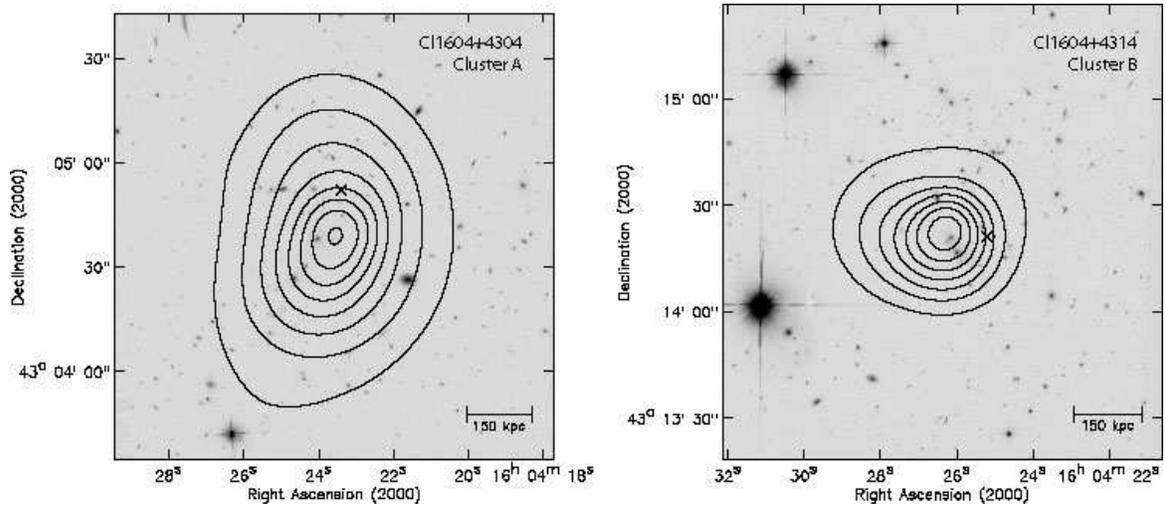}
\caption{Optical imaging from HST of the two Cl1604 clusters (A and
  B) detected in our Chandra imaging with contours of their diffuse X-ray
  emission overlaid.  The contour levels begin at 1.25 times the local
  background and are logarithmically spaced thereafter.  The optical
  centroids of each cluster, derived as the median position of the cluster's confirmed
  members, are marked by the crosses in each panel.  The optical images shown
  are in the F814W filter and have a pixel scale of 0.03 $^{\prime\prime}$/pixel. \label{fig-opt+xray}}
\end{figure*}

\begin{figure}
\epsscale{1.0}
\plotone{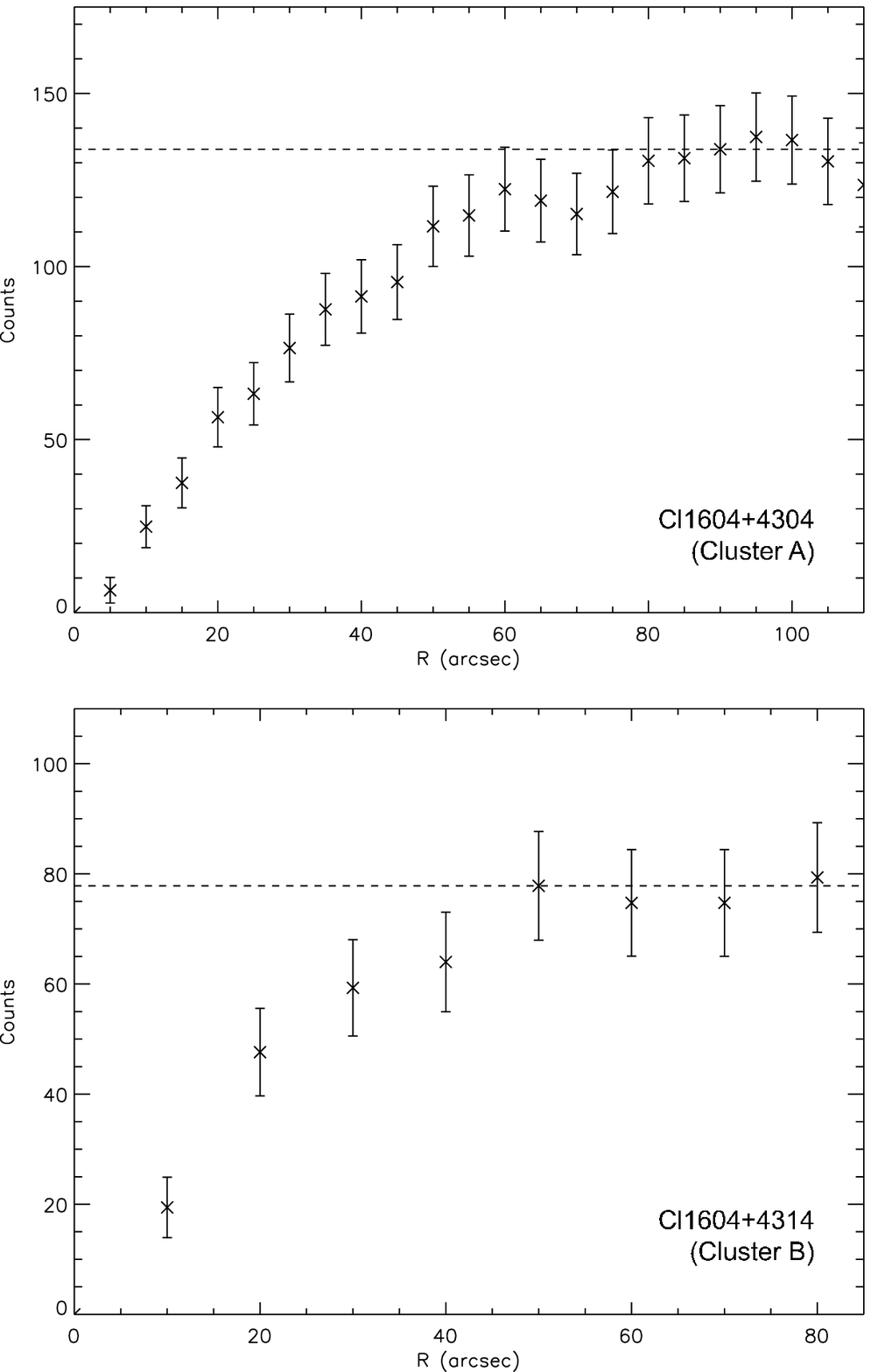}
\caption{Cumulative net count profiles for clusters A and B.  The total
  number of counts originating from each cluster is shown by the horizontal dashed line. \label{fig-gca-clA}}
\end{figure}

A total of 10 red galaxy overdensities were detected in our LFC imaging of the
Cl1604 field (Gal et al.~2008), of which 8 fall within the FOV of our ACIS-I imaging.  The
optical properties and velocity dispersions of these galaxy clusters and
groups are discussed in detail in Gal et al.~(2005, 2008).  We searched for
diffuse X-ray emission at the location of each overdensity and found
significant ($>3\sigma$) emission above the background near clusters
Clusters A and B.  An increased X-ray background in the soft band was
detected near Cluster D and Cl1605+4322 (hereafter Cluster F), but at a much reduced significance level.
For Cluster C, Cl1604+4324 (Cluster G), and Cl1604+4322 (Cluster H), we find
no emission in excess of the background.  The galaxy overdensity Cl1604+4314B
(system E), which is likely a superposition of supercluster members at
various redshifts (see Gal et al. 2008), is also not detected as expected.  The
diffuse emission detected from the Cl1604 systems is shown in Figure
\ref{fig-mosaic_asmoothed_nops}, which displays an adaptively smoothed,
soft-band image of the supercluster field where we have replaced all point sources
detected by {\tt wavdetect} with an estimate of the local background using
{\tt dmfilth} prior to smoothing.  Contours of the X-ray emission from
Clusters A and B are also plotted overtop \emph{HST}-ACS imaging of the
systems in the F814W-band in Figure \ref{fig-opt+xray}.  

To quantify the extent, count rate and significance above the background of
the emission from Clusters A and B we employed a growth curve analysis on the
soft-band counts from each cluster.  Azimuthally averaged surface brightness
profiles were constructed for each system by summing the counts in annuli
centered on the peak of the diffuse emission.  The background level of each
field was then set to the median of the source-free, outer portions of the
profile.  A cumulative net count profile was then constructed by measuring
the counts in successively larger apertures centered on the diffuse emission
and subtracting an appropriately scaled background.  We take the total number
of counts originating from the cluster to be the level at which the
cumulative profile ceases to grow.  The cumulative net count profiles for the
two clusters are shown in Figure \ref{fig-gca-clA}.   We detect a total of
133.9 and 76.3 net counts\footnote{The quoted values are vignetting corrected
  counts above the background} in the soft band within an extent of
$80^{\prime\prime}$ and $50^{\prime\prime}$ (0.62 and 0.39 $h_{70}^{-1}$ Mpc)
from Clusters A and B at a significance level of $7.20\sigma$ and
$6.21\sigma$, respectively.  

\begin{figure}
\epsscale{1.0}
\plotone{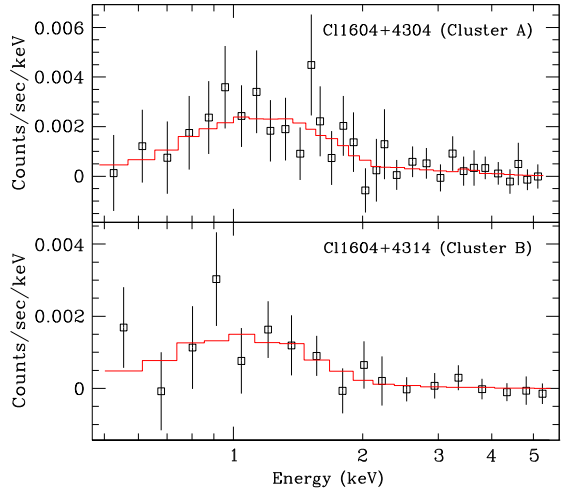}
\caption{Spectral fits to the observed energy spectra of Clusters A and B.
  The spectra have been background subtracted and grouped to contain at least
  15 counts per bin.  The best fit Raymond-Smith thermal spectra have
  temperatures of 3.5 and 1.6 keV for Cluster A and B, respectively. \label{fig-sherpa}}
\end{figure}

Despite the low number counts, we determined the temperature and X-ray
luminosity of each system by fitting a Raymond-Smith thermal plasma model to
the energy spectra of the diffuse emission.  The spectra were extracted out
to the measured extent of each system with the CIAO task {\tt specextract}
and the background was measured in a local annulus surrounding the extraction
region.  The spectra were grouped to contain at least 15 counts per bin and
fit with the {\tt Sherpa} package using a $\chi^{2}$ statistic with the
Gehrels (1986) approximation of errors given the low number counts.  The
metal abundance was fixed at 0.3Z$_{\odot}$ and only the temperatures of the
systems were allowed to float.  The fit was carried out over 0.3-8 keV, but
we found our results to be robust against variations to this energy range.
Our extracted spectra and resulting best-fits are shown in Figure
\ref{fig-sherpa}.  We find temperatures of $3.50^{+1.82}_{-1.08}$ and
$1.64^{+0.65}_{-0.45}$ keV for Clusters A and B, respectively; the estimated
confidence intervals are $1\sigma$ and were derived through projection of the
statistic surface using the proj task in {\tt Sherpa}.  Cluster A was
previously observed with \emph{XMM-Newton} and we find that our best-fit
temperature for the system is higher than the $2.51^{+1.05}_{-0.69}$ keV
reported by Lubin et al.~(2004), but consistent given the errors on both
estimates.  The resulting bolometric luminosities for Cluster A and B are
$1.43\times10^{44}$ and $8.20\times10^{43}$ $h_{70}^{-2}$ erg s$^{-1}$,
respectively, within an extent of $80^{\prime\prime}$ and
$50^{\prime\prime}$.  If we assume the cluster surface-brightness profiles
follow a $\beta$-model with the canonical parameters $\beta=2/3$ and
$r_{c}=180$ $h_{70}^{-1}$ kpc, the measurement apertures enclose 90.9\% and
70.5\% of the total flux out to the $R_{200}$ radius of each cluster. We
determined $R_{200}$ as $R_{200} = 2\sigma_{v} / \sqrt{200} H(z)$ using the
velocity dispersions measured by Gal et al.~(2008); the resulting radii are
$110^{\prime\prime}$ and $126^{\prime\prime}$.  Extrapolating the observed
emission out to $R_{200}$, Clusters A and B have bolometric luminosities of
$1.58\times10^{44}$ and $1.16\times10^{44}$ $h_{70}^{-2}$ erg s$^{-1}$,
respectively.  Our best-fit temperatures and the resulting fluxes and
luminosities of the two clusters are summarized in Table \ref{tab-xprop}.

The remaining five clusters show no statistically significant emission in
excess of the background and are therefore treated as non-detections.  We
calculated upper limits to the counts from each system as the $3\sigma$
Poissonian fluctuation of the background measured within a 0.5 $h_{70}^{-1}$
Mpc radius aperture centered on the system's optical centroid from Gal et
al. (2008).  While we do observe a slightly increased background at the
positions of Clusters D and F, the measured counts are less than that
expected from a $3\sigma$ Poissonian fluctuation, therefore we treat both
systems as non-detections.  The $3\sigma$ upper limit to the counts from each
cluster is listed in Table \ref{tab-xprop}.  We converted count rates to flux by normalizing a Raymond-Smith thermal plasma model in
{\tt Sherpa} to the measured upper limits for each system, with the appropriate
instrument response files obtained using {\tt specextract} at the location of
each cluster.  We assumed a $0.3Z_{\odot}$ metallicity and a temperature of 2
keV for all systems.  The resulting upper limits to the flux and luminosities
of each system are listed in Table \ref{tab-xprop}.  We note that had we
iteratively solved for the cluster temperatures using the $L_{\rm x}-T$
relationship as opposed to using a fixed 2 keV value, the resulting
temperatures would have ranged from 2.4-2.6 keV.  The change would reduce our
luminosity upper limits by less than 1\%.

\subsection{Cluster Scaling Relations}

\begin{figure}[t]
\epsscale{1.0}
\plotone{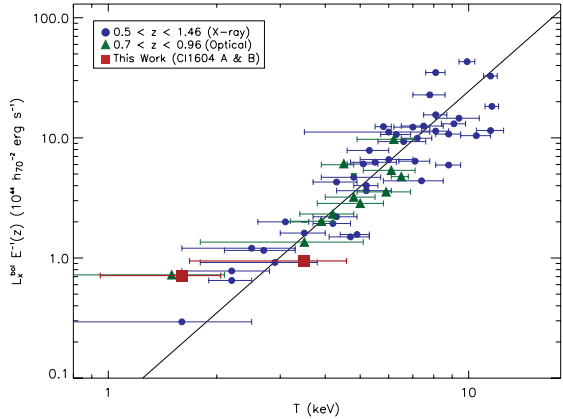}
\caption{Relation between bolometric luminosity and ICM temperature for
  clusters at $z\ge0.5$ scaled to account for self-similar evolution.  Filled
  squares indicate the two optically selected clusters detected in the Cl1604
  Supercluster, while the circles and triangles denote X-ray and optically
  selected high-redshift clusters, respectively, drawn from the literature.  The solid line
  represents the best-fit of Markevitch (1998) to the correlation observed in
  low-redshift clusters. \label{scalrel-LxT}}
\end{figure}

It was previously reported by Lubin et al.~(2004, hereafter L04) that Cluster A and
the optically selected cluster Cl1324+3011 at $z=0.76$ were found to be underluminous
compared to their high-redshift, X-ray selected counterparts with similar galaxy
velocity dispersions.  The systems were also found to deviate from the
$\sigma_{v}-T$ relationship at the $\sim4\sigma$ level, exhibiting significantly cooler ICM
temperatures than expected from local scaling relationships.
These observations hinted at the intriguing possibility that optical cluster
selection at high redshift may preferentially select younger systems that
have yet to to assemble a significant ICM or undergo a major phase of non-gravitational heating.
In this section we revisit this issue using the results of our Chandra
observations and newly determined velocity dispersions for the Cl1604 systems
from Gal et al.~(2008).  

\begin{center}  \emph{The $L_{\rm x}-T$ Relation}  \end{center}

If the thermodynamics of a cluster's ICM is governed solely by gravitational
processes, the self-similar description of clusters predicts X-ray luminosity should scale
with ICM temperature as $L\propto T^{2}$, given the gas radiates via
bremsstrahlung emission (Kaiser 1986).  Furthermore, the evolution of this relationship
is expected to follow the evolution of the Hubble parameter which goes as:

\begin{equation}
E(z) = [\Omega_{m} (1+z)^{3} + (1-\Omega_{m} - \Omega_{\Lambda})(1+z)^{2} +
\Omega_{\Lambda}]^{1/2}
\end{equation}

In actuality, several studies have shown that clusters exhibit hotter ICM temperatures
than expected from the self-similar relationship, suggesting
non-gravitational processes such as AGN feedback have injected energy into
the systems.  Low-redshift cluster surveys have found the $L_{\rm x}-T$
relationship follows a form closer to $L\propto T^{3}$ (Markevitch 1998; Xue \& Wu 2000; Vikhlinin et al.~2002).
On the other hand, although clusters do not obey the predicted $L_{\rm x}-T$  relationship, studies show the
evolution of the correlation does indeed follow the expected self-similar evolution.
Maughan et al.~(2006) have shown that when self-similar evolution is taken
into account, the properties of X-ray selected WARPS clusters out
to $z\sim1$ agree with the low-redshift $L_{\rm x}-T$ relationship, indicating
the processes which heat the ICM beyond the self-similar prediction occur at
an even earlier epoch or in dynamically younger clusters (see also Vikhlinin
et al.~2002 and Hicks et al.~2006).

\begin{figure}[t]
\epsscale{1.0}
\plotone{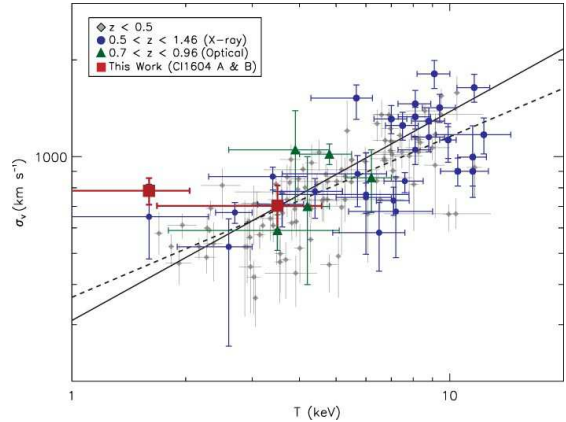}
\caption{Relation between galaxy velocity dispersion and ICM
  temperature.  Filled squares indicate the two optically selected clusters detected in the Cl1604
  Supercluster, while the circles and triangles denote X-ray and optically
  selected, high-redshift clusters, respectively, drawn from the literature.  Diamonds
  indicate the $z<0.5$ sample from Horner et al.~(2001).  The dashed
  line is the best-fit relation of Horner et al.~(2001) to the low redshift
  cluster data, while the solid line is that of Xue \& Wu (2000). \label{scalrel-ST}}
\end{figure}

In Figure \ref{scalrel-LxT} we plot the bolometric X-ray luminosity of
Clusters A and B extrapolated out to their $R_{200}$ radii against their ICM
temperatures.  Also shown are other high-redshift, X-ray and optically
selected clusters drawn from the literature.  We have corrected each system
for the predicted self-similar evolution and overplotted the local $L_{\rm x}-T$
relationship of Markevitch (1998), which has a power-law slope of 2.64.
The optically selected clusters are drawn from the ESO Distant Cluster Survey
(EDiscS; White et al.~2005, Johnson et al.~2006) and the Red Sequence
Cluster Survey (RCS; Gladders \& Yee 2000, Hicks et al.~2007), while the X-ray
selected sample includes clusters from the Massive Cluster Survey (MACS; Ebeling et
al. 2001, 2007) and several individual clusters drawn from a variety of
studies (Borgani et al.~1999; Donahue et al.~1999; Gioia et al.~1999; Tran et
al. 1999; Ebeling et al.~2001; Holden et al.~2001; Stanford et al.~2001, 2002; Vikhlinin et
al. 2002; Valtchanov et al.~2004; Maughan et al.~2004; Rosati et al.~2004; Mullis et al.~2005; Hilton et al.~2007; Demarco
et al.~2007).

In agreement with the results of L04 we find that the optically selected 
Clusters A and B are consistent with the $L_{\rm x}-T$ relationship followed by
high-redshift, X-ray selected systems.  The same is largely true for the
optically selected EDisCs and RCS clusters, as previously reported.  While we
find Cluster B deviates from the best-fit relationship of Markevitch (1998)
at the $2.0\sigma$ level, it is well within the cluster-to-cluster scatter  
observed in the correlation.  Figure  \ref{scalrel-LxT} suggests that
while optical selection of clusters at high redshift often finds clusters
further down the luminosity (and hence mass) function, it does not
preferentially select systems that deviate from the predicted scaling
relationships between cluster X-ray properties.

\begin{figure}[t]
\epsscale{1.0}
\plotone{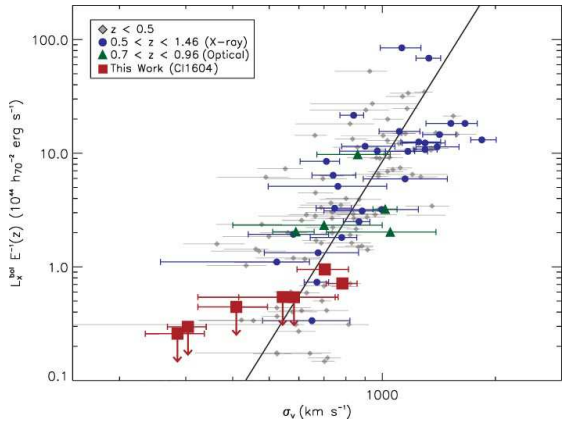}
\caption{Relation between bolometric luminosity and galaxy velocity
  dispersion.  Filled squares indicate the two optically selected clusters detected in the Cl1604
  Supercluster, while the circles and triangles denote X-ray and optically
  selected, high-redshift clusters, respectively, drawn from the literature.  Diamonds
  indicate the $z<0.5$ sample from Horner et al.~(2001).  The solid line
  represents the best-fit relation of Xue \& Wu (2000) to the correlation
  observed in low-redshift clusters.  \label{scalrel-LxS}}
\end{figure}

\begin{center}  \emph{The $\sigma_{v}-T$ Relation}  \end{center}

L04 found that while the X-ray properties of their optically
selected clusters were consistent with X-ray-X-ray scaling relations, they strongly deviated from X-ray-optical
relations, such as the $\sigma_{v}-T$ correlation.  If the gas which makes up the
ICM shares the same dynamics as the cluster galaxies, it is expected that the
ICM temperature should be related to galaxy velocity dispersion as
$\sigma_{v} \propto T^{1/2}$.  L04 reported that their
optically selected clusters had cooler temperatures than expected given
their measured velocity dispersions.   The deviations were at the 4 and 5
$\sigma$ levels for Cl1324+3011 and Cluster A, respectively.  We revisit this issue with an improved
velocity dispersion for Cluster A and the first ever X-ray observations of Cluster B in the Cl1604 system.

Since the publication of the L04 results, our spectroscopic dataset for
Cluster A has substantially improved (see Gal et al.~2008), allowing us to
update the original estimate of its velocity dispersion.  Using a
$3\sigma$ iterative clipping technique on 35 redshifts within 1 $h_{70}^{-1}$
Mpc of the cluster center, we measure a velocity dispersion of $703\pm110$ km
s$^{-1}$ for the system.  Our revised value is significantly lower than the
L04 estimate of 1226 km s$^{-1}$ determined from a sample of 22 galaxies
within a $2^{\prime}\times 8^{\prime}$ region centered on the cluster.  Using
the same procedure on  galaxies in Cluster B, we obtain a velocity dispersion
of $780\pm74$ km s$^{-1}$ for the system.

In Figure \ref{scalrel-ST} we plot the galaxy velocity dispersion vs ICM
temperature of the two detected Cl1604 clusters, as well as several
high-redshift X-ray and optically selected clusters drawn from the
literature.  Also shown in the background are clusters from a large sample of
273 low to moderate redshift systems observed with \emph{ASCA} and uniformly analyzed
by Horner (2001).  Their best fit $\sigma_{v}-T$ relationship using this dataset
is plotted as the dashed line.  Alternatively the solid line shows the best
fit from Xue \& Wu (2000), who used a slightly larger sample of clusters
drawn from the literature and observed using a variety of instruments.

Using our revised velocity dispersion for Cluster A we find
the system is now in very good agreement with the $\sigma_{v}-T$ relation found at low
redshifts and followed by high-redshift, X-ray selected systems.  The same is
largely true for the two EDisCs and three RCS clusters which have published
velocity dispersions. The exception to this is Cluster B which deviates
significantly from both the Xue \& Wu (2000) and Horner (2001) best-fit
relations.  The deviation is significant at roughly the $4\sigma$ level
(without taking into account the observed scatter).

As our spectral fits for Cluster B found it highly unlikely that the system's
temperature is significantly greater than 2 keV, which would place the system
squarely in-line with the $L_{\rm x}-T$ relation, an alternative explanation
for the observed deviation is that the system is not fully relaxed.  In that
case the assumption that the ICM shares the same dynamics as the cluster
galaxies may not hold true.  Gal et al.~(2008) note that the redshift
distribution for Cluster B does shows evidence of velocity segregation
indicative of either substructure or a triaxial system.  They also find that
the velocity dispersions of the red vs blue galaxy populations in the cluster
differ by less than $1\sigma$, unlike the $3.7\sigma$ difference observed in
Cluster A.  The latter is the most isolated cluster in the Cl1604
supercluster with the most prominent red sequence.  If the system formed at
an earlier epoch than Cluster B, it is expected that the primordial red
galaxy population would have had more time to fully virialize and establish a
much different dispersion than any infalling blue galaxy population.  The
lack of a significant difference in the velocity dispersions of blue and red
galaxies in Cluster B may be further evidence that the system is undergoing
collapse or possible merger processes, which may drive the cluster off the $\sigma_{v}-T$ relationship.

\begin{center}  \emph{The $L_{\rm x}-\sigma_{v}$ Relation}  \end{center}

In addition to the findings of L04, Fang et al.~(2007) have reported that their optically-selected clusters
detected in the DEEP2 Galaxy Redshift Survey (Davis et al.~2003) are underluminous relative to their
measured velocity dispersions.  It is expected that the luminosity of the ICM
should scale with galaxy velocity dispersion as $L_{\rm x} \propto \sigma_{v}^{4}$ if
both the gas and galaxies are in virial equilibrium and if the gas mass is
proportional to the virial mass of the system (Quintana \& Melnick 1982).  The L04
and Fang et al.~(2007) results seem to suggest optical selection of cluster may
preferentially select young systems that have not built-up a significant ICM
and are therefore underluminous compared to X-ray selected systems.

In Figure \ref{scalrel-LxS} we plot the bolometric luminosity of the two
detected Cl1604 clusters against their measured velocity dispersions.  We have
also plotted the upper limits obtained for each system in the supercluster
that went undetected, but for which we have measured velocity dispersions of galaxies
within 1 $h_{70}^{-1}$ Mpc of the cluster centers.  Also shown are
several high-redshift X-ray and optically selected clusters drawn from the
literature which have published velocity dispersions.  Having corrected for the
expected self-similar evolution, we compare the high redshift observations to
the best-fit local relationship of Xue \& Wu (2000) and the low redshift
sample of Horner (2001).  Using our revised velocity dispersion for Cluster A, we find the system
is fully consistent with the $L_{\rm x}-\sigma_{v}$ relation.  The same is
true for the clusters which we did not detect in our observations given the
upper limits on their X-ray luminosities. Cluster B exhibits the greatest deviation (at the $2\sigma$ level), but given
the observed scatter about the best-fit, we conclude the system is consistent
with the relationship.

From our results and those of other optically-selected cluster surveys
summarized in Figure \ref{scalrel-LxS}, we conclude that we do not observe a systematic deviation of
optically-selected clusters from the $L_{\rm x}-\sigma_{v}$ relation. 
Regarding the deviation observed by Fang et al.~(2007), we simply point out that the systems that deviated
most significantly from the $L_{\rm x}-\sigma_{v}$ relation were also the systems
whose velocity dispersions were determined using as few as three member
galaxies. Given the challenges of measuring accurate velocity dispersions at
high redshift with dozens of redshifts (see discussion in Gal et al.~2008), we feel
these clusters require further observations before it can be conclusively
determined whether they fall significantly off the $L_{\rm x}-\sigma_{v}$ relationship.

\section{Conclusions}

We have presented the results of \emph{Chandra} observations of the Cl1604 supercluster at
$z=0.9$, the largest such structure mapped at redshifts approaching unity,
with the most constituent clusters and groups and the largest number of
spectroscopically confirmed member galaxies.  Over nearly two ACIS-I
pointings we find a $2.5\sigma$ excess of X-ray point sources in the
hard$_{10}$ band, while no such overdensity is observed in the
soft band.  At a flux of $1\times10^{-14}$ erg s$^{-1}$
cm$^{-2}$ (2-10 keV), the surface density of hard sources is 1.47 times
greater than that of a blank field, in excellent agreement with the
correlation between cluster redshift and source overdensity observed by
Cappelluti et al. (2005).  Unlike many previous reports, the overdensity
persists when integrated over nearly two full ACIS-I pointings, making it
unlikely the excess is solely due to statistical fluctuations in the source
counts.   If the excess sources are tracing substructure within the Cl1604
system, then this observation supports the notion that more obscured, harder
X-ray sources are more highly biased tracers of large-scale structure than their softer counterparts. 

Using a maximum likelihood technique we have matched 112 of the 161 detected X-ray point
sources to optical counterparts found in our Palomar 5m-LFC and
\emph{HST}-ACS imaging, of which 42 have spectroscopic information available. 
We find 15 sources that are associated with the supercluster, all of which
have rest-frame luminosities consistent with emission from AGN activity.
The supercluster AGN largely avoid the densest regions of the system and are
instead distributed on the outskirts of massive clusters or within poorer
clusters and groups.  We find a large fraction of the AGN in or near Cluster C.  The
system has a modest velocity dispersion ($\sigma = 386$ kms $^{-1}$) and
we do not detect diffuse emission from the system in these observations.
Despite the high density of galaxies in Cluster A and extensive spectroscopic
coverage of the system, we find no AGN in or near the cluster.  On the other
hand sources are found on the outskirts of Cluster B, which has an equally
high velocity dispersion and a luminous ICM.  The primary difference between
the two clusters noted by Gal et al. (2008) is that Cluster A appears fully
relaxed, while Cluster B shows signs of velocity substructure indicative of a
recent merger or a significant population of actively accreting galaxies.  It
is possible that the more complex dynamical state of Cluster B leads to
increased galaxy interactions and/or mergers on the outskirts of the system
which may trigger enhanced AGN activity.  

We have detected diffuse emission from Clusters A and B, while
the remaining five clusters and groups in the supercluster show no
significant emission above the background.  We find that Clusters A and B have
bolometric luminosities of $1.43\times10^{44}$ and $8.20\times10^{43}$
$h_{70}^{-2}$ erg s$^{-1}$ and gas temperatures $3.50^{+1.82}_{-1.08}$ and
$1.64^{+0.65}_{-0.45}$ keV, respectively.  Using updated velocity dispersions
from Gal et al.~(2008), we find that the properties of Cluster A agree well
with both X-ray-X-ray and X-ray-optical cluster scaling relations followed by
high-redshift, X-ray selected galaxy clusters.  On the other hand we find
Clusters B deviates from the $\sigma_{v}-T$ relationship at the $\sim4\sigma$ level.
This may be due to the system's complex dynamical state, indicating the
cluster is not fully relaxed and may still be in the process of forming.  

\acknowledgments
We thank Patrick Henry and Harald Ebeling for several useful
discussions. This work is supported by the Chandra General Observing Program
under award number GO6-7114X. Additional support for this program was
provided by NASA through a grant HST-GO-11003 from the Space Telescope
Science Institute, which is operated by the Association of Universities for
Research in Astronomy, Inc., under NASA contract NAS 5-26555. The
spectrographic data used herein were obtained at the W.M. Keck Observatory,
which is operated as a scientific partnership among the California Institute
of Technology, the University of California and the National Aeronautics and
Space Administration. The Observatory was made possible by the generous
financial support of the W.M. Keck Foundation.

\clearpage

\LongTables
\begin{landscape}
\tabletypesize{\scriptsize}
\setlength{\headsep}{2in}

\end{center}

\end{document}